\documentclass[sigconf]{acmart}
\makeatletter
\renewcommand\@formatdoi[1]{\ignorespaces}
\makeatother
\renewcommand\footnotetextcopyrightpermission[1]{} 
\newcommand\scalemath[2]{\scalebox{#1}{\mbox{\ensuremath{\displaystyle #2}}}}
\usepackage{ amssymb }
\usepackage{multirow}

\newtheorem{theorem}{Theorem}

\makeatother

\newcommand{\bx}{\mbox{\bf x}}

\newcommand{\corr}{\mathrm{corr}}

\begin{document}

\title{Dealing With Ratio Metrics in A/B Testing at the Presence of Intra-User Correlation and Segments}

%
\author{Keyu Nie}
\authornote{Both authors contributed equally to this research.}
\email{knie@ebay.com}
\affiliation{%
  \institution{eBay Inc}
}
\author{Yinfei Kong}
\authornotemark[1]
\email{yikong@fullerton.edu}
\affiliation{%
  \institution{California State University Fullerton}
}

\author{Ted Tao Yuan}
\email{teyuan@ebay.com}
\affiliation{%
  \institution{eBay Inc}
  }

\author{Pauline Berry Burke}
\email{pmburke10@gmail.com}
\affiliation{%
  \institution{eBay Inc}
}

\renewcommand{\shortauthors}{Keyu and Yinfei, et al.}

\begin{abstract}
We study ratio metrics in A/B testing at the presence of correlation among observations coming from the same user and provides practical guidance especially when two metrics contradict each other. We propose new estimating methods to quantitatively measure the intra-user correlation (within segments). With the accurately estimated correlation, a uniformly minimum-variance unbiased estimator of the population mean, called correlation-adjusted mean, is proposed to account for such correlation structure. It is proved theoretically and numerically better than the other two unbiased estimators, naive mean and normalized mean (averaging within users first and then across users). The correlation-adjusted mean method is unbiased and has reduced variance so it gains additional power. Several simulation studies are designed to show the estimation accuracy of the correlation structure, effectiveness in reducing variance, and capability of obtaining more power. An application to the eBay data is conducted to conclude this paper.
\end{abstract}


\maketitle 
\thispagestyle{empty}
\section{Introduction}
The A/B testing is an online randomized controlled experiment that compares the performance of a new design B (treatment) of web service with the current design A (control). It randomly assigns traffic/experiment units (users, browse id/guid, XID and so on) into one of the two groups, collects metrics of interests, conducts hypothesis testing to claim significant treatment effect and estimates the average treatment effect/lift (ATE) over the whole targeting group. In this paper, we would call a random experiment unit as a user. Statistically, the simple A/B test is a two-sample $t$-test but there are many situations that the data structure is no longer simply two independent groups of observations with independent and identical distribution (i.i.d.). For example, a user may visit a website multiple times so those visits can be considered as repeated measures of the same user. These repeated measures are identical but not independent, as there should be a certain level of correlation among repeated measures within the user. Such intra-user correlation has big impacts when dealing with ratio metrics, where ratio metrics involve the analysis units having smaller granular level comparing with experiment units, such as click-through rate (CTR), average selling price (ASP), search result page to view item (SRP to Vi), search result page exit rate (SRP Exit Rate) and so on. 

At the same time, there can be various types of users such as cell phone users, laptop users, and desktop users. In this paper, we call a type of user as a segment. Those users in the same segment shall share some level of behavioral similarities, such as independent and identical distribution. Thus, we consider a three-level data structure, segment-user-observation, as follows. Observations within a user are identically and commonly correlated. However, observations from different users but the same segment shall be identically and independently distributed. We further assume that the distribution of all measures (i.e., observations) from the same segment, whichever users they are from, is identical. Across different segments, such distribution could be different in means. The main purpose of this article is to shed light on ratio metrics in A/B test at the presence of repeated measures of users as well as multiple segments of users.

Many researchers have considered the same kind of data structure \citep{yandexratio, deng2018}, and delta method is suggested to estimate the variance of naive mean estimators. Here naive mean is to take the average of all repeated measures, and it is common to use the difference of naive means as an unbiased estimator of ATE. The supporters \citep{yandexratio, deng2018} argue that it naturally matches the definition of the metrics, like CTR as the summation of all clicks divided by the summation of all impressions, taking the expected number of repeated measures as part of the metrics. On the other hand, we notice that there is another way of computing this ratio metrics called normalized mean: compute the ratio metrics for each user/experiment-unit and then take the average of these user-level normalized metrics. Researchers \citep{yandexuser} observed that the two ratio metrics could lead to contradicted conclusions via a two-sample $t$-test. However, in Kohavi et al.'s new book \citep{kohavi2020trustworthy}, he discussed two estimators (naive/normalized mean) and argued that both are useful definitions. Normalized mean is generally recommended in \citep{kohavi2020trustworthy}, as it is ``more robust to outliers, such as bots having many page views or clicking often". More other recent theoretical investigation of the variance of ratio metrics can be found in \citep{Sekhon2020}.

Despite the various opinions among different researchers in this area, this article provides another view of this problem and proposes practical guidance especially when two metrics contradicted each other. We offer new estimating methods to quantitatively measure the intra-user correlation (within segments). With the accurately estimated correlation, we can construct a uniformly minimum-variance unbiased estimator (UMVUE) of the population mean called ``correlation-adjusted mean" for the data with repeated measures. Hence, we shall have improved power in the A/B test. We formulate the problem in detail in Section~\ref{formulate}, deploy the proposed method in Section~\ref{approach}, and support it with numerical analysis in Section~\ref{numeric}.

\section{Problem Formulation} \label{formulate}
Suppose we are conducting a randomized controlled experiment and denote each experiment unit by $i$. We observe repeated measurements of metrics $(X_{i1},\dots,X_{in_i})$ where $n_i$ denotes the number of repetitions for unit $i$. This data structure is used in this paper to describe ratio metrics. In the context of A/B testing for website comparison, an experiment unit is a user while repeated measures refer to the multiple website visits from the same user. We are interested in testing if the treatment effect is significant by the means of $\mathrm{E}(X)$ per event, like click through rate (CTR) or average sold price (ASP). Naturally there are two ratio metrics describing this \citep{kohavi2020trustworthy}:
\begin{align*}
\bar{R}^{A} = \frac{\sum_{i=1}^{N}\sum_{j=1}^{n_i} X_{ij}}{\sum_{i=1}^{N} n_i}, \quad \bar{R}^{B} = N^{-1} \sum_{i=1}^{N} \frac{\sum_{j=1}^{n_i} X_{ij}}{n_i}.
\end{align*}
Both ratios can be used to estimate the population mean but which one is better. What are the differences between the two? From many people, $\bar{R}^{A}$ is more natural comparing with $\bar{R}^{B}$ and many papers from Linkedin, Microsoft \citep{deng2018}, Uber \citep{zhao2018safely}, Yandex \citep{yandexratio}, \citep{yandexuser} follow this definition. Yandex \citep{yandexuser} reported that these two metrics can have different directional indications. 
In this paper, we want to inline with \citep{kohavi2020trustworthy} and argue both metrics are natural estimates of the population mean but they are different in many aspects. For notation convenience, we call $\bar{R}^{A}$ as naive mean and $\bar{R}^{B}$ as normalized mean.

\section{Approach} \label{approach}
We consider the problem of choosing between these two ratios in a randomized controlled experiment with repeated measures under special designs as described above. Under this structure, instead of assuming i.i.d. of $X_{ij}$, we should expect the repeated measures share a common mean with constant intra-user correlation.The $X_{ij}$ are identical distributed with correlation $\rho$ and follow the same distribution $\textbf{\textit{F}}$ for each unit/user $i$.
Let $E(X_{ij}) = \mu$ and $Var(X_{ij}) =\sigma^2$. It is easy to show that $r_i = \sum_{j=1}^{n_i} X_{ij}/n_i$ is an unbiased estimator of $\mu$ with variances $\sigma^2_{r_i}=\sigma^2\{(1-
\rho)/n_i+\rho\}$ {(please see Appendix \ref{apen2} for proof)}. 
\begin{theorem} \label{thm1}
With the model setting above, we define
\begin{align*}
S_1 &=\frac{1}{\sum_{i=1}^{N} n_i -1 }    \sum_{i=1}^{N} \sum_{j=1}^{n_i}(X_{ij} - \bar{R}^{A} )^2, \\ S_2 &=\frac{1}{\sum_{i=1}^{N} n_i -1 }    \sum_{i=1}^{N} \sum_{j=1}^{n_i}(X_{ij} - \bar{R}^{B} )^2, \\
S_3 &= \frac{1}{\sum_{i=1}^{N} (n_i -1) } \sum_{i=1}^{N} \sum_{j=1}^{n_i}\sum_{j^{'}=1 \; and \; j^{'} \neq j}^{n_i}(X_{ij} -r_i)(X_{ij^{'}} -r_i).
\end{align*}
Then we have
\begin{align*}
E(S_1) & \approx \sigma^2, \quad
 E(S_2) \approx \sigma^2, \quad
E(S_3)  = ( \rho - 1 ) \sigma^2, \\
\end{align*} 

\end{theorem} 
The proof of Theorem \ref{thm1} can be found in the Appendix \ref{apen1}. We can get the estimates of $\rho$ and $\sigma^2$ from above.
{Review that in the literature of univariate repeated measures, it follows the designs as $Y_{ij}=\mu+\pi_i+\tau_j+e_{ij}$ with $\pi_i \sim N(0, \sigma_{\pi}^2)$ representing the individual difference component for observation $i$, $\tau_j \sim N(0, \sigma_{\tau}^2)$ representing random deviation due to repeated measures, and $e_{ij} \sim N(0, \sigma_{e}^2)$ representing error. With such formulation, $\corr(Y_{ij}, Y_{ij'}) = \sigma_{\pi}^2 / (\sigma_{\pi}^2 + \sigma_{\tau}^2)$ is a parameter independent of $i$ or $j$. This correlation is usually called intra-class correlation. Although our data design does not necessarily follow this formulation, we argue that it is reasonable to assume $\corr(Y_{ij}, Y_{ij'})$ is a constant for the problem of repeated measures among users.} 

\subsection{Unified View of Two Ratio Metrics} \label{chp3.1}
Suppose we have a total of $N$ units/users. Then for an arbitrary weight vector $W=(w_1,\dots,w_N)^T$ with $\sum_{i=1}^{N}w_i=1$, the following quantity is also an unbiased estimator of $\mu$: 
\begin{align*}
\bar{R}^{W} = \sum_{i=1}^{N} w_i r_i.
\end{align*}
With such definition, we see that if $W=(1/N,\dots,1/N)^T$, then $\bar{R}^{W} = \bar{R}^{B}$. It is also clear that if $W=(n_1/\sum_{i=1}^{N} n_i,\dots,n_N/\sum_{i=1}^{N} n_i)^T$, then $\bar{R}^{W} = \bar{R}^{A} $.
Both $\bar{R}^{A}$ and $\bar{R}^{B}$ are special cases and unbiased estimators of $\mu$ under repeated measures with common mean data structure. The difference is that $\bar{R}^{A}$ (naive mean) utilizes weights proportional to repeated measure count $n_i$ over $r_i$ per unit, but $\bar{R}^{B}$ (normalized mean) equally weights $r_i$ from each unit. 

In fact, the uniformly minimum-variance unbiased estimator (UMVUE) \citep{cochran1937} of $\mu$ (proof in Appendix \ref{apen4}) is by setting
\begin{equation}
w_i \propto \frac{1}{\sigma^2_{r_i}}  \propto \frac{n_i}{1+(n_i -1)\rho}. \label{set1}
\end{equation}

Theorem \ref{thm1} provides a quantitative way to measure the intra-user correlation as $\hat{\rho}$. Based on the magnitude of $\hat{\rho}$, we could compare the variance between naive mean ($\bar{R}^{A}$) and normalized mean($\bar{R}^{B}$). 
Specifically, when $\rho=1$, $\bar{R}^{B}$ is the UMVUE of $\mu$, hence $Var(\bar{R}^{B}) < Var(\bar{R}^{A})$. Similarly when $\rho=0$ we have $Var(\bar{R}^{A}) < Var(\bar{R}^{B})$. 
We suggest following the rule based on our observation: \
if $\hat{\rho} < \frac{\frac{\sum_{i=1}^{N} \frac{1}{n_i}}{N^2} - \frac{1}{\sum_{i=1}^{N} n_i} }{\frac{\sum_{i=1}^{N} n_i^2}{(\sum_{i=1}^{N}n_i)^2} + \frac{\sum_{i=1}^{N} \frac{1}{n_i}}{N^2} - \frac{1}{\sum_{i=1}^{N} n_i} - \frac{1}{N}},$ then $Var(\bar{R}^{A}) < Var(\bar{R}^{B})$; Otherwise, $Var(\bar{R}^{B}) < Var(\bar{R}^{A})$.

\medskip
\noindent \textit{Remarks.} 
{(1) Normalized mean($\bar{R}^{B}$) is a fairness metric, since it treats each user equally. In the meanwhile, naive mean($\bar{R}^{A}$) could easily attribute the treatment changes (in percent) to numerator and denominator. Both ratio metrics are useful in final reports. In practice, $\bar{R}^{A}$ is slightly preferred in search-related experiments and $\bar{R}^{B}$ is preferred in advertisement experiments.}

{(2) As mentioned in \citep{deng2018}, delta method should be adapted to estimate the variance of the naive mean($\bar{R}^{A}$). However, we could directly utilize sample variance to estimate the variance of the normalized mean($\bar{R}^{B}$).}

\subsection{Optimized $\bar{R}^{W}$ under UMVUE} \label{chp3.2}

It is shown in equation \ref{set1} that the optimal (UMVUE) weight $w_i = \frac{n_i}{1+(n_i -1)\rho} \approx n_i^{1-\rho}$ (in Appendix \ref{apen3}), where $n_i^{1-\rho}$ represents the effective sample size. The optimized $\bar{R}^{W}$ (UMVUE of $\mu$) would be:
\begin{align} \label{umvue}
\bar{R}^{\rho} = \sum_{i=1}^{N} \frac{n_i^{1-\rho}}{\sum_{i=1}^{N} n_i^{1-\rho} } r_i.
\end{align}
We denote it as correlation-adjusted mean. By plugging in the estimator of $\rho$ in Theorem \ref{thm1}, this correlation-adjusted mean would have the smallest variance. Of course, Delta method should be applied to estimate its variance. We are primarily focusing on Improving metric sensitivity (aka. variance reduction) in our large scale trustworthy experimentation platform. The option proposed here follow this guideline: we always prefer metric with smaller variance.    


%
%

\subsection{What to Do When Two Ratio Metrics Have Contradiction} \label{simpsonSec}
Both the naive mean and normalized mean are unbiased estimators of the common mean $\mu$. Ideally, we should expect the two matches to each other in direction. In practice, it is not always like that. Actually, at eBay, we also found similar cases mentioned in \citep{yandexuser} that the signs of treatment lift from naive mean and normalized mean are contradicted with each other. We utilize a simplified example with Simpson's paradox to illustrate the contradictory conclusions can even be possible using these two ratio metrics. A randomized controlled experiment is conducted in which two observations are repeatedly measured multiple times. Table \ref{tab1} summarizes the experiment results.
\begin{table}
\centering
\begin{tabular}{lllll} %
\toprule
\multirow{2}{*}{}                          & \multicolumn{2}{l}{\textcolor[rgb]{0.133,0.133,0.133}{Control} }           & \multicolumn{2}{l}{Treatment}                                           \\ 
\cmidrule(lr{.75em}){2-3}\cmidrule(lr{.75em}){4-5}
                                           & $\sum_{j=1}^{n_i} X_{ij}$ & $n_i$ & $\sum_{j=1}^{m_i} Y_{ij}$ & $m_i$  \\ 
\midrule
\textcolor[rgb]{0.133,0.133,0.133}{Obs. 1 (Segment 1)} & 200                                                              & 300      & 20                                                           & 24        \\ 
\midrule
\textcolor[rgb]{0.133,0.133,0.133}{Obs. 2 (Segment 2)} & 10                                                               & 30       & 100                                                          & 200       \\
\bottomrule
\end{tabular} 
\caption{Simpson's Paradox} \label{tab1}
\end{table}
We calculate the two ratio metrics:
\begin{align*}
\bar{R}^{A}_{ctr} &= \frac{200+10}{300+30} =\frac{7}{11}, &\quad \bar{R}^{A}_{trt} = \frac{20+100}{24+200} =\frac{15}{28}, \\
\bar{R}^{B}_{ctr} &= \frac{1}{2}\left(\frac{200}{300} + \frac{10}{30}\right) =\frac{1}{2}, &\quad \bar{R}^{B}_{trt} = \frac{1}{2}\left(\frac{20}{24} + \frac{100}{200}\right) =\frac{2}{3}
\end{align*}
where the sub-indices $ctr$ and $trt$ are introduced to denote the control and treatment group respectively. Therefore, the treatment effect can be estimated as:
\begin{align*}
\hat{\Delta}^{A} = \bar{R}^{A}_{trt} - \bar{R}^{A}_{ctr} =-0.10, \quad \hat{\Delta}^{B} = \bar{R}^{B}_{trt} - \bar{R}^{B}_{ctr} =0.17.
\end{align*}
Since the signs of this two estimates are different, the conclusion of comparison between treatment and control is inconsistent. 
The example explains the contradictory of the two ratio metrics potentially accounts for Simpson's paradox. 

Usually, Simpson's paradox happens when there are heterogeneous treatment effects, or the ratio-metric $R$ has differences in segments. The contradicted conclusion from naive mean and normalized mean would be a sign of Simpson's paradox. Further analysis in each user segment separately is usually recommended for this situation, we would discuss it in more details in Section \ref{chp3.4}. Continuously monitoring differences of sample count ($N$) and repeated measure count ($\sum_{i=1}^{N}n_i$) from test and control in user segments will help us alert such situation (two metrics with contradiction) in the early stage. 

\subsection{Generalization to multiple user segments} \label{chp3.4}

In the context of website comparison, most of the time, experiment units, i.e. users, could be from different segments. Common user segments include device type (desktop vs mobile), browser type, country, and so on. Metrics in different user segments usually are different. For instance, it is reasonable to assume desktop users have different CTR from mobile users, as they may experience different UI designs. To avoid Simpson's paradox, as well as possible contradicted signs, it is always better to analyze it in segment view, and follow the guidance in Sections \ref{chp3.1} and \ref{chp3.2}. 

To integrate the estimators from multiple segments, we can use the weighted sum of these estimators from segments with the segment weight proportional to either the sample size of each segment (weight on users) or the repeated measure size of each segment (weight on the repeated measure). The choice of segment weights is beyond the scope of this article but we refer to \citep{freedman2008regression, linBerkeley, miratrix2013} for more discussion on it. We can show that naive mean ($\bar{R}^{A}$) of whole users is the same as integrating multiple naive means of each segment with segment-weights on the number of repeated measure count of each segment; normalized mean ($\bar{R}^{B}$) integrate multiple normalized means of each segment with segment-weights on the number of users (sample count) of each segment.

\section{Numerical Analysis} \label{numeric}

We illustrate the effectiveness of the proposed methods with several simulation studies and real data analysis. {In the simulation studies, we present four examples to demonstrate the following points.

First, we want to evaluate the accuracy of our proposed estimator (in Theorem \ref{thm1}) for the intra-user correlation $\rho$ and show that our estimator is accurate. Second, We investigate how the standard errors of $\bar{R}^{A}$ and $\bar{R}^{B}$ vary with $\rho$.  We will show that the true standard deviation of $\bar{R}^{\rho}$ is smaller than that of $\bar{R}^{A}$ and $\bar{R}^{B}$. We estimate $\rho$ from Theorem 1 and plug it in Equation (\ref{umvue}) to obtain the correlation-adjusted mean. We empirically calculate the true standard deviation of $\bar{R}^{\rho}$, $\bar{R}^{A}$, and $\bar{R}^{B}$ with bootstrapping (1000 iterations). Third, we demonstrate that more power can be obtained in the A/B test by combining the proposed estimator $\bar{R}^{\rho}$ obtained from different segments in treatment and control groups. We simulation an additional example that mimics the real A/B testing where the test group and control group are both comprised of users coming from different segments.  Finally, in the fourth example, we consider a more general correlation structure instead of a common correlation as assumed in our method. But we will see that the correlation adjustment idea still works in the general situation. It is also implying the robustness of common correlation assumption.

In addition to the simulation studies, we apply it to an eBay real data set. 
We find that the intra-user correlation $\rho$ is presented in our data set, and provide estimates of the correlations for different metrics. We also conclude that the proposed estimator $\bar{R}^{\rho}$ does provide a more sensitive metric to measure the success of an A/B test. To simplify the process and highlight the scope of the paper, we only evaluate the standard errors of three different estimators within one experiment group (treatment or control) using bootstrap. 

As we explained above that we design four simulation examples for four different purposes. The first example only contains one segment of users and we want to show the estimation accuracy of $\rho$ as below.}  

\medskip
\textbf{Example 1}. We simulate $1000$ users in the following manner. For the $i$-th user, we generate the number of observations $n_i$ from $\text{Poisson}(10) + 1$. Denote each observation by $X_{i,j}$ with $j \in \{1, \cdots, n_i\}$. Let $X_{i,j} \sim \text{Bernoulli}(p_i)$ with $p_i \sim N(0.3, 0.05)$. Note that observations belonging to the same user may not necessarily independent of each other. We set $\text{corr}(X_{ij}, X_{ij'}) = \rho$ for $j \ne j'$. There are a number of ways to introduce correlation structure for Bernoulli distribution. We choose to firstly generate multivariate normal observations $\widetilde{\bx}_{i\cdot} = (\widetilde{X}_{i1}, \cdots, \widetilde{X}_{in_i})^T \sim N(\textbf{0}, \Sigma(\widetilde{\rho}))$ with diagonal elements of $\Sigma(\widetilde{\rho})$ being $1$ and off-diagonal elements being $\widetilde{\rho}$. We then dichotomize $\widetilde{X}_{ij}$ to $\{0, 1\}$ at $\Phi^{-1}(1-p_i)$ where $\Phi(\cdot)$ is the cumulative density function of a standard normal distribution. In other words, we let $X_{ij} = 1$ if $\widetilde{X}_{ij} > \Phi^{-1}(1-p_i)$ and $X_{ij} = 0$ otherwise. We consider $\widetilde{\rho} \in \{0, 0.2, 0.4, 0.6, 0.8\}$ and present the estimation results in columns 3-5 of Table \ref{sim1}. The column 3 presents the true correlation between dichotomized $X_{i,j}$ and $X_{ij'}$. Note that the new correlation ($\rho$) is different from their correlation ($\widetilde{\rho}$) before dichotomization. The experiments are repeated $1000$ times.

\begin{table}[]
\centering
\scalebox{1}{
\begin{tabular}{lllllllllll}
\toprule
       $\widetilde{\rho}$\quad \quad \quad & $\rho$\quad \quad\quad \quad   & Method\quad \quad & $\hat{\rho}$\quad\quad\quad\quad & $\text{SD}(\hat{\rho})$ \\
\midrule
 0      & 0.0022      &   S3/S1            & 0.0117      & 0.0050            \\
 0.2    & 0.1204     &  S3/S1            & 0.1292      & 0.0093          \\
 0.4    & 0.2488    &  S3/S1            & 0.2559      & 0.0126            \\
 0.6    & 0.3947    &   S3/S1            & 0.4015      & 0.0150            \\
 0.8    & 0.5773     &   S3/S1            & 0.5813      & 0.0161        \\
\cmidrule(lr{.75em}){1-5} 
 0      & 0.0022        &   S3/S2           & 0.0117      & 0.0050         \\
 0.2    & 0.1204       &   S3/S2         & 0.1292      & 0.0093           \\
 0.4    & 0.2488        &   S3/S2         & 0.2559      & 0.0126            \\
 0.6    & 0.3947      &   S3/S2         & 0.4015      & 0.0150         \\
 0.8    & 0.5773        &   S3/S2         & 0.5813      & 0.0161         \\
\bottomrule
\end{tabular}
}
\caption{Estimation Accuracy of $\rho$ in Example 1}\label{sim1}
\end{table}

The estimation accuracy of $\rho$ for Example 1 is presented in Table \ref{sim1}. The S3/S1 and S3/S2 refer to the estimation methods of $\rho$ by calculating the sample version of $E(S_3)/E(S_1) + 1$ and $E(S_3)/E(S_2) + 1$, respectively. The columns $\hat{\rho}$ and $\text{SD}(\hat{\rho})$ denotes the mean and standard deviation of estimated values based on $1000$ repetitions. We see from the Table \ref{sim1} that our methods estimate $\rho$ very accurately since $\hat{\rho}$ is fairly close to the true value $\rho$.

\medskip
\textbf{Example 2}. The capability of accurately estimating $\rho$ is shown in Example 1. We now proceed to show how the change of intra-user correlation $\rho$ interacts with the variances of naive mean ($\bar{R}^{A}$) and normalized mean ($\bar{R}^{B}$). We also show the optimized estimator of $\mu$, $\bar{R}^{\rho}$, will have a reduced variance compared to others. We want to show it for more values of $\rho$ and show the robustness of our method under various settings. Therefore, we simulate a separate example here. Following the notation in Example 1, we generate $1000$ users with $n_i \sim \text{Poisson}(2) + 1$, $p_i \sim N(0.3, 0.04)$, and $\rho \in \{0.1, 0.2, \cdots, 0.9\}$.

We then present the results of Example 2 in Table \ref{variance_red}. We only choose S3/S1 for illustration purposes in Example 2. Results should be similar if the other estimator is used. In Table \ref{variance_red}, column $\bar{y}$ gives the true value of $\mu$; column $\hat{y}$ gives the estimated value of $\mu$; column $\hat{\sigma}_{\bar{y}}$ gives the estimated standard deviation for each method calculated based on values of $1000$ repetitions. Three methods, naive mean ($\bar{R}^{A}$), normalized mean ($\bar{R}^{B}$), and the proposed method, weighted mean ($\bar{R}^{\rho}$) are included in the table. It is clear that the correlation-adjusted mean ($\bar{R}^{\rho}$) has the smallest standard deviation across all values of $\rho$. In the meanwhile, the standard error of $\bar{R}^{A}$ is smaller than $\bar{R}^{B}$ if $\rho < 0.3$, and the standard error of $\bar{R}^{A}$ is bigger than $\bar{R}^{B}$ if $\rho > 0.3$.

\begin{table}[]
\centering
\scalebox{0.8}{
\begin{tabular}{llllll}
\toprule
        & $\bar{y}$ (truth) \quad\quad& $\hat{y}$ (estimate)\quad\quad & $\hat{\sigma}_{\bar{y}}$ (estimate)\quad\quad & $\rho$ \\
\cmidrule(lr{.75em}){2-5}
Naive        & 0.3   & 0.29971       & 0.00947       & 0.1 \\
Normalized \quad\quad & 0.3   & 0.29953      & 0.01016       & 0.1 \\ 
Corr. Adj.                  & 0.3   & 0.29972      & 0.00942      & 0.1 \\
\cmidrule(lr{.75em}){2-5}
Naive        & 0.3   & 0.30006      & 0.01041       & 0.2  \\
Normalized  & 0.3   & 0.29997      & 0.01074       & 0.2  \\
Corr. Adj.                  & 0.3   & 0.30007     & 0.01026       & 0.2  \\
\cmidrule(lr{.75em}){2-5}
Naive        & 0.3   & 0.29968      & 0.01126       & 0.3  \\
Normalized  & 0.3  & 0.29947      & 0.01126       & 0.3  \\
Corr. Adj.                  & 0.3   & 0.29966      & 0.01100       & 0.3  \\
\cmidrule(lr{.75em}){2-5}
Naive        & 0.3   & 0.30015      & 0.01206       & 0.4  \\
Normalized  & 0.3   & 0.30009      & 0.01177       & 0.4  \\
Corr. Adj.                  & 0.3   & 0.30016      & 0.01160       & 0.4  \\
\cmidrule(lr{.75em}){2-5}
Naive        & 0.3   & 0.30020      & 0.01280       & 0.5 \\
Normalized  & 0.3   & 0.30005      & 0.01226       & 0.5 \\
Corr. Adj.                  & 0.3   & 0.30015      & 0.01219       & 0.5 \\            
\cmidrule(lr{.75em}){2-5}
Naive        & 0.3   & 0.29998      & 0.01349       & 0.6 \\
Normalized  & 0.3   & 0.29965      & 0.01274       & 0.6 \\
Corr. Adj.                  & 0.3   & 0.29982      & 0.01270       & 0.6 \\
\cmidrule(lr{.75em}){2-5}
Naive        & 0.3   & 0.30064      & 0.01417       & 0.7 \\
Normalized  & 0.3   & 0.30060      & 0.01321       & 0.7 \\
Corr. Adj.                  & 0.3   & 0.30062      & 0.01321       & 0.7 \\
\cmidrule(lr{.75em}){2-5}
Naive        & 0.3   & 0.29952      & 0.01479       & 0.8 \\
Normalized  & 0.3   & 0.29953      & 0.01364       & 0.8 \\
Corr. Adj.                  & 0.3   & 0.29955      & 0.01362       & 0.8 \\
\cmidrule(lr{.75em}){2-5}
Naive        & 0.3   & 0.29999      & 0.01540       & 0.9 \\
Normalized  & 0.3   & 0.29974      & 0.01406       & 0.9 \\
Corr. Adj.                 & 0.3   & 0.29977      & 0.01407       & 0.9\\
\bottomrule
\end{tabular}
}
\caption{Variance reduction compared to naive and normalized mean in Example 2}\label{variance_red}
\end{table}

\medskip
\textbf{Example 3}. We further illustrate that additional power can be gained if using the proposed weight mean method. We design an example where users in the treatment and control groups come from different segments. Let us focus on the control group first. Consider three segments of subjects/users following the multinomial distribution $(C_1, C_2, C_3)^T \sim \text{Multi-nomial}(1/3, 1/2, 1/6)$. For segment 1, $n_i \sim \text{Poisson}(2) + 1$ and $X_{i,j} \sim \text{Bernoulli}(p_i)$ with $p_i \sim N(0.3, 0.04)$. For segment 2, $n_i \sim \text{Poisson}(5) + 1$ and $X_{i,j} \sim \text{Bernoulli}(p_i)$ with $p_i \sim N(0.5, 0.08)$. For segment 3, $n_i \sim \text{Poisson}(30) + 1$ and $X_{i,j} \sim \text{Bernoulli}(p_i)$ with $p_i \sim N(0.7, 0.04)$. In any of the three segments, we set $\text{corr}(X_{ij}, X_{ij'}) = 0.3$. Then we simulate the data for the treatment group in a very similar way but the only difference is $p_i$ follows $N(0.3+d, 0.04)$, $N(0.5+d, 0.08)$, and $N(0.7+d, 0.04)$, respectively. We consider $d \in \{0.01, 0.02, \cdots, 0.08\}$. The experiment is repeated $1000$ times for each $d$. {Figure 1 shows the comparison of power in this example.

\begin{figure}[!htbp] \centering
\begin{center}%
\begin{tabular}
[l]{l}%
{\hspace{-0.1in}\includegraphics[scale=0.3]{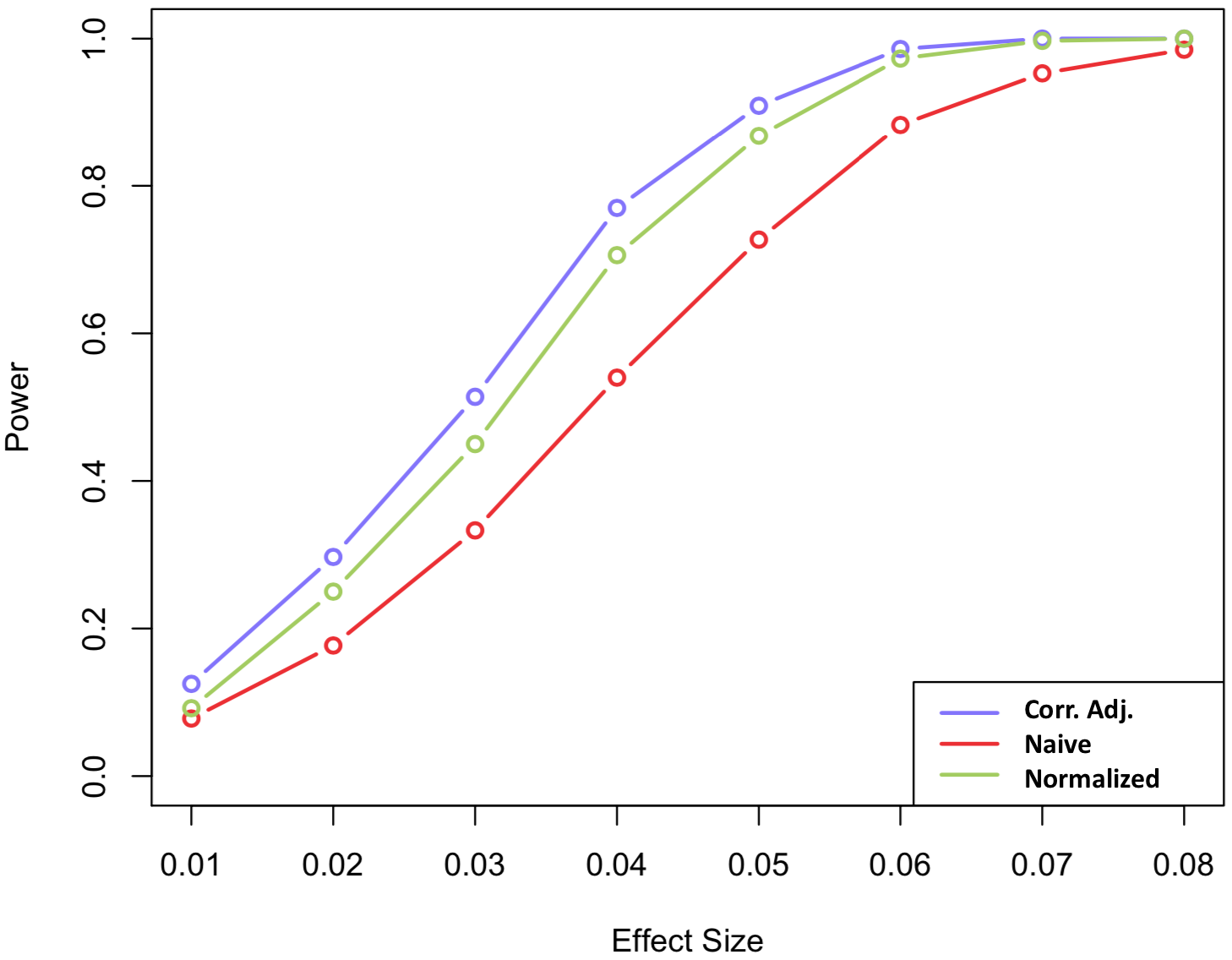}
}
\end{tabular}
\vspace{-0.2in}
\caption{Power analysis in Example 3}
\label{Fig1}%
\end{center}%
\end{figure}%

\medskip
\textbf{Example 4}. We explore the effectiveness of the proposed method in handling the unequal correlation structure. To be more specific, we borrow the setting in example 2 but the correlation between $X_{ij}$ and $X_{ij'}$ is $\rho^{|j - j'|}$ for any $j \ne j'$. We set $\rho = 0.9$ as it is common to have a high correlation for repeated measures next to each other but low correlation when they are far from each other. Note that the true correlation structure is no longer equal between observations but in a more general autoregressive pattern.

The results for this example are presented in Table \ref{variance_ex4}. The standard deviation is smaller based on our correlation adjusted method. It also suggests that assuming constant intra-user correlation is robust when the true data structure violates the assumption}

\begin{table}[]
\centering
\scalebox{0.8}{
\begin{tabular}{llllll}
\toprule
               & $\bar{y}$ (truth) \quad\quad& $\hat{y}$ (estimate) \quad\quad& $\hat{\sigma}_{\bar{y}}$ (estimate)  \\
\cmidrule(lr{.75em}){2-4}
Naive        & 0.3 & 0.28557 & 0.01394     \\
Normalized \quad\quad & 0.3 & 0.29205 & 0.01355      \\
Corr. Adj.                 & 0.3 & 0.29074 & 0.01342      \\
\bottomrule
\end{tabular}
}
\caption{Variance reduction compared to naive and normalized mean in Example 4}\label{variance_ex4}
\end{table}

\subsection{Validation in eBay Data}
We extend our validation into eBay real data to verify how the significance of $\hat{\rho}$ guides the relevant variance compare between naive mean and normalized mean, as well as
how much improvement (with correlation-adjusted mean) can be achieved. We randomly sampled search activities from users with each size of 500,000 users on eBay global site with primary metrics as { ``Ratio \textbf{$\mathit{R}_1$}'' (alike Exit Rate) for UK and ``Ratio \textbf{$\mathit{R}_2$}'' (alike CTR) for US } in some treatment groups. We further repeated 5 replicas of the previous process { with equal size 500,000} and showed the average estimation results at table \ref{realData}.  { The reported metric estimations were added a constant to hide the real information, but it should not change our conclusion in follows.} To properly evaluate the estimator standard errors, we use bootstrap (1000 iterations) with re-sampling id being the experiment unit id to be an educated guess at each replica. 

Our results in Table \ref{realData} compared the standard errors of naive mean ($\bar{R}^{A}$), normalized mean ($\bar{R}^{B}$) and correlation-adjusted mean ($\bar{R}^{\rho}$). The method S3/S1 is used to accurately estimate the intra-user correlation $\rho$. It clearly revealed that the intra-user correlation coefficient $\rho$ did present in our data set, and varied from $9.6\%$ to $56\%$ for ``Ratio \textbf{$\mathit{R}_1$}" at UK site and ``Ratio \textbf{$\mathit{R}_2$}" at US site respectively. {The root cause of different intra-user correlated behaviors ($\rho$) between the UK and US is actually due to a search feature launched in the US site only. The dependence among repeated measures within users could not be ignored and should be carefully addressed to pick the right estimation method.   For small $\rho$ $(9.6\%)$, $\bar{R}^{A}$ showed $3.3\%$ improvement on standard error comparing with $\bar{R}^{B}$, which in hence saved $1-(1-3.3\%)^2=6.6\%$ on sample sizes. On the contrast, for big $\rho$ $(56\%)$, $\bar{R}^{B}$ showed $16.7\%$ improvement on sample sizes comparing with $\bar{R}^{A}$. For both cases,  $\bar{R}^{\rho}$ saved up to $15.2\%$ and $40.3\%$ sample sizes correspondingly.

We noticed the bias between the naive mean and normalized mean for ``Ratio \textbf{$\mathit{R}_2$}" could be a sign of heterogeneous user effects presence. A proper segment classification algorithm is needed to further analyze this heterogeneous effect, which is beyond the scope of this paper.

\begin{table}[]
\centering
\scalebox{0.7}{
\begin{tabular}{llllll}
\toprule
Method            \quad\quad\quad\quad\quad\quad             & Metric Name  \quad\quad & Site \quad\quad & Estimate  \quad\quad\quad & Std. Error \quad\quad & $\hat{\rho}$ \\
\midrule
Naive ($\bar{R}^{A}$)      & Ratio \textbf{$\mathit{R}_1$} & UK   & 0.42850048 & 0.00061878 & 0.09618527   \\
Normalized ($\bar{R}^{B}$) & Ratio \textbf{$\mathit{R}_1$} & UK   & 0.43028952 & 0.00064033 & 0.09618527   \\ 
Corr. Adj. ( $\bar{R}^{\rho}$)                      & Ratio \textbf{$\mathit{R}_1$} & UK   & 0.42839652 & 0.00058949 & 0.09618527   \\ 
Naive ($\bar{R}^{A}$)      & Ratio \textbf{$\mathit{R}_2$}     & US   & 0.49476930  & 0.00129048 & 0.56069220    \\ 
Normalized ($\bar{R}^{B}$) & Ratio \textbf{$\mathit{R}_2$}     & US   & 0.52730064 & 0.00107419 & 0.56069220    \\ 
Corr. Adj.  ($\bar{R}^{\rho}$)                         & Ratio \textbf{$\mathit{R}_2$}     & US   & 0.51028800   & 0.00077074 & 0.56069220    \\
\bottomrule
\end{tabular}
}
\caption{Estimation average of 5 replicas with each sample size = 500,000 users from eBay data} \label{realData}
\end{table}

\section{Conclusion and Restriction}
In this paper, we studied how to estimate ratio metrics in a randomized controlled experiment with repeated measures within each experiment unit. As the intra-user correlation coefficient $\rho$ could not be ignored, we established a way to accurately measure the severity of intra-user correlation coefficients $\rho$. We showed the naive mean ($\bar{R}^{A}$) and normalized mean ($\bar{R}^{B}$) were both weighted user means with different weights. We proved there is no clear winner between naive mean and normalized mean when considering estimator variance, as it highly depends on the severity of intra-unit correlation coefficients $\rho$. We further proposed a correlation-adjusted mean ($\bar{R}^{\rho}$), which adopted the optimal weights depending on $\rho$. Our simulation and real data empirical analysis validated that we can accurately estimate $\rho$ and built more sensitive ratio-metric estimators based on $\rho$. Due to the variance reduction technique, we shall improve the power as well as requiring less sample size for A/B testing, which improves experiment efficiency.

The main restriction of the proposed method is on its assumption of a common correlation for different users within the same segment. It is possible that there is a variation of such correlation for the same type of users (from the same segment). Such observation motivates us to consider more flexible settings where the correlation for a user within the same segment may follow a random distribution, and the random distribution can be different across segments. However, this type of setting is the subject of our future work.


\bigskip
\bigskip
\bigskip

\appendix
{
\section{Variance Formula of Sample Mean with Correlated Measure} \label{apen2}
\textbf{Claim}:  The $X_{j}$ are identical distributed with correlation $\rho$ and follow the same distribution $\textbf{\textit{F}}$. Let $E(X_{j}) = \mu$ and $Var(X_{j}) =\sigma^2$. Then:
\begin{itemize}
\item[a.] $r = \sum_{j=1}^{n} X_{j}/n$ is an unbiased estimator of $\mu$.
\item[b.] The variance of $r$ is $\sigma^2_{r}=\sigma^2\{(1-\rho)/n+\rho\}$. 
\end{itemize} 
\begin{proof}
Since \begin{align*}
E(r) &= E(\sum_{j=1}^{n} X_{j}/n) = \sum_{j=1}^{n}E(X_{j})/n = \sum_{j=1}^{n}\mu/n = \mu,
\end{align*}
then $r = \sum_{j=1}^{n} X_{j}/n$ is an unbiased estimator of $\mu$.

The variance of $r$ is
\begin{equation*}
\scalemath{0.85}{
\begin{aligned}
\sigma^2_{r} &= E(r - E(r))^2 = E(r - \mu)^2 = E(\sum_{j=1}^{n} X_{j}/n - \mu)^2 = \frac{1}{n^2} E(\sum_{j=1}^{n} (X_{j} - \mu) )^2 \\                      
                      &= \frac{1}{n^2}( \sum_{j=1}^{n}E(X_{j} - \mu)^2 + \sum_{k=1}^{n} \sum_{m=1, m \neq k}^{n} E(X_{k} - \mu)(X_{m} - \mu) ) \\                 
                      &= \frac{1}{n^2}( (\sum_{j=1}^{n}\sigma^2 + \sum_{k=1}^{n} \sum_{m=1, m \neq k}^{n} \sigma^2 \rho ) =  \frac{1}{n}\sigma^2 + \frac{n-1}{n} \sigma^2 \rho     \\
                      &=  \sigma^2\{(1-\rho)/n+\rho\}.
\end{aligned}
}
\end{equation*}
\end{proof}
}
\section{Proof of UMVUE} \label{apen4}
\textbf{Claim}: Suppose $r_i$ with $i = 1, \dots, N$ are independent with each other and $E(r_i) = \mu$, $Var(r_{i}) =\sigma^2_{r_{i}}$. Then 
\begin{itemize}
\item[a.] for an arbitrary weight vector $W=(w_1,\dots,w_N)^T$ with $\sum_{i=1}^{N}w_i=1$, the following quantity is also an unbiased estimator of $\mu$: $\bar{R}^{W} = \sum_{i=1}^{N} w_i r_i$. 
\item[b.] the UMVUE of $\mu$ is by setting: $w_i \propto \frac{1}{\sigma^2_{r_i}}$. 
\end{itemize}
\begin{proof}
\begin{align*}
E(\bar{R}^{W}) &= E(\sum_{i=1}^{N} w_i r_i) = \sum_{i=1}^{N} w_i E(r_i) = \sum_{i=1}^{N} w_i \mu = \mu.
\end{align*}
Thus, $\bar{R}^{W}$ is an unbiased estimator of $\mu$.

To find the minimum value of $Var(\bar{R}^{W}) = \sum_{i=1}^{N} w_i^{2} \sigma^2_{r_i}$ under restriction $\sum_{i=1}^{N}w_i=1$, we use method of Lagrange multiplier: $\mathcal{L} = \sum_{i=1}^{N} w_i^{2} \sigma^2_{r_i} - \lambda (\sum_{i=1}^{N}w_i - 1)$.
Since $\frac{\partial^2 \mathcal{L}}{\partial w_i^{2}} = 2 \sigma^2_{r_i} > 0$, so $\mathcal{L}$ is a concave function of $w_i$. The minimum value of $Var(\bar{R}^{W})$ is equal to the minimum of $\mathcal{L}$, and it achieves at:
\begin{align*}
\frac{\partial \mathcal{L}}{\partial w_i} &=  2 w_i \sigma^2_{r_i} - \lambda w_i= 0 \, for \, i =  1, \dots, N,\\
 \frac{\partial \mathcal{L}}{\partial \lambda} &= \sum_{i=1}^{N}w_i - 1 = 0. 
\end{align*}
By solving equations above we have $w_i = \frac{\frac{1}{\sigma^2_{r_i}}}{\sum_{i=1}^{N}\frac{1}{\sigma^2_{r_i}}} \propto \frac{1}{\sigma^2_{r_i}}$.
\end{proof}
\section{Approximation} \label{apen3}
\textbf{Claim}:  $\frac{n}{1+(n -1)\rho} \approx n^{1-\rho}$.
\begin{proof}
\

From Taylor series expansion, the 1st degree polynomial approximation of function $f(n) = n^{\rho}$ at $n = 1$ is $n^{\rho} \approx  1 + (n - 1) \rho$.
Therefore,
\begin{align*}
n^{1 - \rho} \approx  \frac{n}{1+(n -1)\rho}.
\end{align*}
\end{proof}

\section{Proof of Theorem \ref{thm1}} \label{apen1}
Without loss of generality, we set $\mu =0$. It is obvious that:
\begin{equation*}
\scalemath{0.85}{
\begin{aligned}
E(r_i^2) &= ( \frac{1}{n_i}  + \frac{n_i -1}{n_i} \rho ) \sigma^2, \\
E(\bar{R}^{A^2}) &= (\frac{1}{\sum_{i=1}^{N}n_i})^2 \sum_{i=1}^{N} n_i^2 E(r_i^2) \notag = (\frac{1}{\sum_{i=1}^{N}n_i})^2 \sum_{i=1}^{N} n_i \{ 1 + (n_i -1) \rho \} \sigma^2 \notag \\
                            &= \left\{\frac{1}{\sum_{i=1}^{N} n_i } + \frac{ \sum_{i=1}^{N} n_i (n_i -1)  }{ ( \sum_{i=1}^{N} n_i )^2 }  \rho \right\} \sigma^2, \\
E(\bar{R}^{B^2}) &= (\frac{1}{\sum_{i=1}^{N}1 })^2 \sum_{i=1}^{N} E(r_i^2) \notag = (\frac{1}{N})^2 \sum_{i=1}^{N} n_i^{-1} \{ 1 + (n_i -1) \rho \} \sigma^2. \\
\end{aligned}
}
\end{equation*}
We also have the following for any $i, j$:
\begin{equation*}
\scalemath{0.85}{
\begin{aligned}
E(r_i X_{i^{'}j}) &= \left\{
\begin{array}{rcl}
0       &      & if \;  i \neq i^{'}\\
( \frac{1}{n_i}  + \frac{n_i -1}{n_i} \rho ) \sigma^2     &      & if \; i = i^{'},
\end{array} \right. \\
E( \bar{R}^{A} X_{ij} ) &= \frac{ n_i E( r_i X_{ij} ) }{\sum_{i=1}^{N}n_i}  \notag =  \frac{ 1 + ( n_i - 1 ) \rho }{\sum_{i=1}^{N} n_i }  \sigma^2, \\
E( \bar{R}^{B} X_{ij} ) &= \frac{1}{N}  E(r_i X_{ij}) \notag =  \frac{ 1 + ( n_i - 1 ) \rho }{N n_i }  \sigma^2.
\end{aligned}
}
\end{equation*}
Thus,
\begin{equation*}
\scalemath{0.85}{
\begin{aligned}
E\{( X_{ij} - \bar{R}^{A} )^2\} & =  E ( X_{ij}^2 - 2 \bar{R}^{A} X_{ij} + \bar{R}^{A^{2}} ) \\
&=  \sigma^2 - 2 \frac{ 1 + ( n_i - 1 ) \rho }{\sum_{i=1}^{N} n_i }  \sigma^2 + (\frac{1}{\sum_{i=1}^{N} n_i } + \frac{ \sum_{i=1}^{N} n_i (n_i -1)  }{ (\sum_{i=1}^{N} n_i)^2 }  \rho ) \sigma^2, \\
E\{( X_{ij} - \bar{R}^{B} )^2\} &=  E ( X_{ij}^2 - 2 \bar{R}^{B} X_{ij} + \bar{R}^{B^{2}} ) \\ 
&=  \sigma^2 - 2 \frac{ 1 + ( n_i - 1 ) \rho }{N n_i }  \sigma^2 + (\frac{1}{N})^2 \sum_{i=1}^{N} n_i^{-1} ( 1 + (n_i -1) \rho ) \sigma^2, \\
E\{ ( X_{ij} -r_i )&( X_{ij^{'}} -r_i ) \} =  E ( X_{ij} X_{ij^{'}}  -  r_i X_{ij} - r_i X_{ij^{'}} + r_i^{2} ) \\
& =  \rho \sigma^2 - ( \frac{1}{n_i}  + \frac{n_i -1}{n_i} \rho ) \sigma^2 =  \frac{1}{n_i}  ( \rho - 1 ) \sigma^2, \\
\end{aligned}
}
\end{equation*}

Therefore, we have
\begin{equation*}
\scalemath{0.7}{
\begin{aligned}
E(S_1) &= \frac{1}{\sum_{i=1}^{N} n_i -1 }    \sum_{i=1}^{N} \sum_{j=1}^{n_i} \left[ \sigma^2 - 2 \frac{ 1 + ( n_i - 1 ) \rho }{\sum_{i=1}^{N} n_i }  \sigma^2 + \left\{\frac{1}{\sum_{i=1}^{N} n_i } + \frac{ \sum_{i=1}^{N} n_i (n_i -1)  }{ ( \sum_{i=1}^{N} n_i )^2 }  \rho \right\} \sigma^2 \right] \\
          &= \frac{\sigma^2}{\sum_{i=1}^{N} n_i -1  } \sum_{i=1}^{N} \left[ n_i -  \frac{ 2 n_i + n_i ( n_i - 1 ) \rho }{\sum_{i=1}^{N} n_i }  + \left\{\frac{ n_i }{\sum_{i=1}^{N} n_i } + n_i \frac{ \sum_{i=1}^{N} n_i (n_i -1)  }{ ( \sum_{i=1}^{N} n_i )^2 }  \rho \right\} \right] \\
          &= \frac{\sigma^2}{\sum_{i=1}^{N} n_i -1  } \left[ \frac{(  \sum_{i=1}^{N} n_i )^2 - \sum_{i=1}^{N} n_i  }{ \sum_{i=1}^{N} n_i }  -  \frac{ \{\sum_{i=1}^{N} n_i (n_i -1) \} \rho }{\sum_{i=1}^{N} n_i }  \right] \\
          &= \sigma^2 \left \{ 1 - \frac{ \sum_{i=1}^{N} n_i (n_i -1)  }{( \sum_{i=1}^{N} n_i ) ( \sum_{i=1}^{N} n_i -1 ) }  \rho\right\} \\
          & \approx \sigma^2,
\end{aligned}
}
\end{equation*}
\begin{equation*}
\scalemath{0.7}{
\begin{aligned}
E(S_2) &= \frac{1}{\sum_{i=1}^{N} n_i -1 }    \sum_{i=1}^{N} \sum_{j=1}^{n_i}\left[ \sigma^2 - 2 \frac{ 1 + ( n_i - 1 ) \rho }{N n_i }  \sigma^2 + (\frac{1}{N})^2 \sum_{i=1}^{N} n_i^{-1} \{ 1 + (n_i -1) \rho \} \sigma^2 \right] \\
          &= \frac{ \sigma^2 }{\sum_{i=1}^{N} n_i -1 } \sum_{i=1}^{N}\left[ n_i -  \frac{ 2 + (n_i -1) \rho }{ N } + n_i (\frac{1}{N})^2 \sum_{i=1}^{N} n_i^{-1} \{ 1 + (n_i -1) \rho \}\right] \\
          &=  \frac{ \sigma^2 }{\sum_{i=1}^{N} n_i -1 } \left[ \left\{\sum_{i=1}^{N} n_i - 2 + \frac{ ( \sum_{i=1}^{N} n_i ) ( \sum_{i=1}^{N} n_i^{-1} ) }{N^2 }\right\} + \frac{  ( \sum_{i=1}^{N} n_i )(N - \sum_{i=1}^{N} n_i^{-1})  - N \sum_{i=1}^{N}(n_i -1) }{ N^2 } \rho \right] \\ 
          &=  \frac{ \sigma^2 }{\sum_{i=1}^{N} n_i -1 } \left[ \left\{\sum_{i=1}^{N} n_i - 2 + \frac{ ( \sum_{i=1}^{N} n_i ) ( \sum_{i=1}^{N} n_i^{-1} ) }{N^2 }\right\} + \frac{  N^2 - ( \sum_{i=1}^{N} n_i )(\sum_{i=1}^{N} n_i^{-1}) }{ N^2 } \rho \right] \\  
          &=  \sigma^2  \left\{ 1 + \frac{  N^2 - ( \sum_{i=1}^{N} n_i )(\sum_{i=1}^{N} n_i^{-1})  }{ N^2 ( \sum_{i=1}^{N} n_i -1 )} ( \rho - 1 ) \right\} \\
          & \approx \sigma^2,
\end{aligned}
}
\end{equation*}
\begin{equation*}
\scalemath{0.7}{
\begin{aligned}
E(S_3) &=  \frac{1}{\sum_{i=1}^{N} (n_i -1) } \sum_{i=1}^{N} \sum_{j=1}^{n_i}\sum_{j^{'}=1 \; and \; j^{'} \neq j}^{n_i}\{\frac{1}{n_i}  ( \rho - 1 ) \sigma^2 \}  =   \frac{1}{\sum_{i=1}^{N} (n_i -1) } \sum_{i=1}^{N} ( n_i - 1 )( \rho - 1 ) \sigma^2 \\
            &= ( \rho - 1 ) \sigma^2,
\end{aligned}
}
\end{equation*}

\end{document}